\documentclass[12pt]{article}
\usepackage{amsfonts,amssymb,epsfig,amsmath}
\usepackage{slashed}
\usepackage{color}
\usepackage{hyperref}
\renewcommand{\text}[1]{#1}

\newcommand{\be}{\begin{equation}}

\newcommand{\ee}{\end{equation}}
\newcommand{\ben}{\begin{displaymath}}
\newcommand{\een}{\end{displaymath}}
\newcommand{\bea}{\begin{eqnarray}}
\newcommand{\eea}{\end{eqnarray}}
\newcommand{\bean}{\begin{eqnarray*}}
\newcommand{\eean}{\end{eqnarray*}}
\newcommand{\nn}{\nonumber \\}
\newcommand{\ba}{\begin{array}}
\newcommand{\ea}{\end{array}}
\newcommand{\bi}{\begin{itemize}}
\newcommand{\ei}{\end{itemize}}

\renewcommand{\theequation}{\arabic{section}.\arabic{equation}}
\def\theequation{\thesection.\arabic{equation}}

\def\l{\lambda}
\def\a{\alpha}

\def\g{\gamma}
\def\G{\Gamma}

\def\G{\Gamma}
\def\g{\gamma}
\def\e{\epsilon}
\def\s{\sigma}
\def\e{\epsilon}

\DeclareMathOperator{\vol}{vol}

\begin{document}

\makeatletter
\renewcommand{\theequation}{\thesection.\arabic{equation}}
\@addtoreset{equation}{section}
\makeatother

\begin{titlepage}

\vfill
\begin{flushright}
FPAUO-12/10\\
\end{flushright}

\vfill

\begin{center}
   \baselineskip=16pt
   {\Large \bf Beyond LLM in M-theory. }
   \vskip 2cm
    Eoin \'O Colg\'ain
       \vskip .6cm
             \begin{small}
      		 \textit{Departamento de F\'isica, 
		 Universidad de Oviedo, \\
 33007 Oviedo, Spain}
             \end{small}\\*[.6cm]
\end{center}

\vfill \begin{center} \textbf{Abstract}\end{center} \begin{quote}
The Lin, Lunin, Maldacena (LLM) ansatz in $D=11$ supports two independent Killing directions when a general Killing spinor ansatz  is considered. Here we show that these directions always commute, identify when the Killing spinors are charged, and show that both their inner product and resulting geometry are governed by two fundamental constants. In particular, setting one constant to zero leads to $AdS_7 \times S^4$, setting the other to zero gives $AdS_4 \times S^7$, while flat spacetime is recovered when both these constants are zero. Furthermore, when the constants are equal, the spacetime is either LLM, or it corresponds to the  Kowalski-Glikman solution where the constants are simply the mass parameter.  
\end{quote} \vfill

\end{titlepage}

%\tableofcontents
%%%%%%%%%%%%%%%%%%%%%%%%%%%%%%%%%%%%%%%%%%%%%%%%%%%%%%%%%%%%%%%%%%%%%%%%%%%%%%%%%%%
%%%%%%%%%%%%%%%%%%%%%%%%%%%%%%%%%%%%%%%%%%%%%%%%%%%%%%%%%%%%%%%%%%%%%%%%%%%%%%%%%%%
\section{Introduction}
Given a supersymmetric solution with an anti-de Sitter ($AdS$) factor in $D=11$ supergravity,  it is expected that a corresponding superconformal field theory (SCFT) exists \cite{Maldacena:1997re}. Beyond the near-horizon of M2 and M5-branes, one way to get more intricate geometries is to consider wrapped branes. In the past, explicit solutions of this nature have been constructed in lower-dimensional gauged supergravities \cite{MN, Acharya:2000mu, Gauntlett:2000ng, Gauntlett:2001qs, Gauntlett:2001jj}, and subsequent efforts have been made directly in higher-dimensions to characterise large classes of $AdS$ geometries by employing wrapped brane ans\"atze \cite{Gauntlett:2006ux, MacConamhna:2006nb,Figueras:2007cn}. While the latter offers greater generality, primarily because one is not confined to a particular dimensional reduction, a recognised advantage of working in lower-dimensions is that it is easier to construct explicit solutions. A notable recent example is a new class of four-dimensional $\mathcal{N}=1$ SCFT \cite{Bah:2011vv,Bah:2012dg} duals generalising \cite{MN}\footnote{See \cite{Kim:2012ek} for a recent discussion on the construction of solutions in higher-dimensions highlighting some associated difficulties.}.  

More generally still, it is possible to discard wrapped brane intuition completely and embrace powerful Killing spinor techniques in higher-dimensions \cite{Gauntlett:2002fz,Gauntlett:2003wb}. In the process the Killing spinor equations (KSE) are converted into differential conditions on spinor bilinears (differential forms) which characterise the spacetime. In $D=11$ this approach has been applied widely \cite{Martelli:2003ki,Kim:2006qu,Kim:2007hv,Colgain:2010wb,Gabella:2012rc}, but two prominent examples concern geometries dual to four-dimensional $\mathcal{N}=1$ \cite{Gauntlett:2004zh} and $\mathcal{N}=2$ SCFTs \cite{LLM}. The beauty here is that the expected $SO(4,2)$ conformal symmetry gets encoded in the $AdS_5$ factor, while the key distinction between $\mathcal{N} =1$ and $\mathcal{N}=2$ depends on the presence of a two-sphere encoding the $SU(2)$ R-symmetry geometrically. Up to warpings, these ans\"atze are extremely general and an added bonus is that, in each case, a $U(1)$ R-symmetry emerges for free form the KSE analysis. 

Quotients of $AdS_5 \times S^5$ aside \cite{Aharony:1998xz}, all known solutions corresponding to $\mathcal{N} =2$ SCFT dual geometries can be traced to the Lin, Lunin, Maldacena (LLM) class of geometries \cite{LLM}, earning it a billing as the most general class of geometries dual to $\mathcal{N}=2$ SCFTs.  The generality of the LLM geometries has recently been strengthened by the observation \cite{genLLM}
that a missing four-form flux in the analysis of LLM  is inconsistent with supersymmetry. Given their uniqueness, the LLM geometries serve as an important basis for subsequent studies. In particular, the gravity duals of a large class of $\mathcal{N}=2$ generalised quiver gauge theories \cite{Gaiotto:2009we} were identified and analysed by Gaiotto and Maldacena \cite{GM}. As discussed in \cite{GM}, since solutions to the LLM class of geometries are in one-to-one correspondence with solutions to the continuum Toda equation, the task of finding solutions can be simplified greatly by the introduction of a global $U(1)$ and reduction to type IIA supergravity, a setting where the Toda equation is replaced by an easier to solve Laplace equation. This simplification has facilitated recent solutions  \cite{ReidEdwards:2010qs,Aharony:2012tz}. 

It is also worth bearing in mind that recent developments in our understanding of $\mathcal{N}=2$ gauge theories mean that geometric insights are not solely confined to supergravity solutions. By using localization, the $S^4$ partition function of $\mathcal{N}=2$ gauge theories can be reduced to a finite-dimensional matrix integral \cite{Pestun:2007rz}, allowing one to study the free energy \cite{Russo:2012ay} and circular Wilson loops  \cite{Rey:2010ry, Passerini:2011fe,Fraser:2011qa,Bourgine:2011ie} at strong coupling  in the large-$N$ limit.  Current findings are in agreement with the suggestion in \cite{Gadde:2009dj} that the dual theory may be sub-critical with only seven ``geometric" dimensions comprising an $AdS_5$ and $S^1$ factor. This appears consistent with a recent search for smooth $AdS_5 \times S^2$ geometries in type IIB supergravity, which concludes that the $SU(2)$ R-symmetry must be non-geometric \cite{Colgain:2011hb}.  

Against this background, in this paper we take the LLM ansatz in $D=11$ to its logical conclusion. Recall that LLM \cite{LLM} initially introduced a general Killing spinor ansatz only to truncate it once a flux term along the internal spacetime was removed. While it was subsequently shown that there are no supersymmetric $AdS_5 \times S^2$ geometries in $D=11$ supported by this omitted flux \cite{genLLM}, an unexpected by-product was the emergence of an extra Killing direction beyond the expected $SU(2) \times U(1)$ R-symmetry. Although LLM can be recovered by identifying the two Killing directions \cite{genLLM}, more generally it is valid to ask if either of them can play a r\^ole as a global $U(1)$ such as in \cite{GM}.  

So, in this paper we tidy up a loose-end in \cite{genLLM}, by treating these two Killing directions independently. Following a review in the next section, in section \ref{sec:commute} we show that the Killing vectors always commute, while their inner product is proportional to the product of two fundamental \textit{constant} scalar bilinears. We calculate the norm of the vectors and observe that when the two fundamental constants are equal that one either has LLM or a spacetime with a null Killing vector. Indeed, it is in this case that we note in \ref{sec:R} that it is always possible to find a linear combination so that the Killing spinors are independent of this direction. Later in section \ref{sec:pp} we confirm that this null spacetime is indeed the maximally supersymmetric pp-wave in $D=11$. 

More generally, when the two constants differ, an extreme case of which is when one of them is zero, we see that the Killing spinors are always charged with respect to both directions indicating that supersymmetry must be double that of LLM. It is then well-known \cite{FigueroaO'Farrill:2002ft} that $AdS_4 \times S^7$, $AdS_7 \times S^4$ and flat spacetime are the only timelike possibilities. A noticeable caveat here is that the Killing spinors are uncharged for flat spacetime, so to fill this gap in the argument, in section \ref{sec:flat} we determine the Killing spinors and the geometry by integrating the differential conditions directly. Our observations here on the existence of different branches of supersymmetric solutions in $D=11$ supergravity mirror findings in related settings, notably half-bps spacetimes with isometry $SO(2,2) \times SO(4) \times SO(4)$ \cite{Lunin:2007ab, D'Hoker:2008wc}. 

In the rest of this paper, case by case, we reduce the Killing spinors of the known solutions from $D=11$ down on $S^5$ and $S^2$ to isolate the two Killing spinors of the LLM ansatz. This allows us to construct all the bilinears explicitly, confirm that the constant bilinears take the expected form and satisfy ourselves that the differential conditions of \cite{genLLM} are satisfied. As we will see in due course, the involved form of these spinors, especially for the Freund-Rubin spacetimes, suggests that using spinor bilinears is not an ideal way to solve the Killing spinor equations, and that these are better solved in $D=11$. On the other hand, through the language of spinor bilinears, we are in a position to make general statements about supersymmetric spacetimes beyond LLM.

\section{Review}
We begin with a lightning review of $D=11$ supergravity solutions preserving $SO(6) \times SO(3)$ symmetry\footnote{As in \cite{LLM, genLLM} we consider the analytically continued geometries  with $S^5$ appearing.}. The $D=11$ supergravity ansatz may be written as a warped product of $S^5$, $S^2$ and a Lorentzian signature  spacetime, $\mathcal{M}_4$,
\bea
\label{ansatz}
ds^{2} &=& e^{2 \lambda} \left[ \frac{1}{m^2} d \Omega_5^2 + e^{2 A} d \Omega_2^2 + ds^2_{\mathcal{M}_4} \right], \nn
F^{(4)} &=& \mathcal{G} + \vol(S^2) \wedge \mathcal{F},
\eea
where the warp factors, $\lambda$ and $A$, are functions of the coordinates on $\mathcal{M}_4$ and $\mathcal{F}$ and $\mathcal{G}$ are respectively two-forms and four-forms on $\mathcal{M}_4$. $m$ is a constant denoting the inverse radius of $S^5$. Throughout this work where $m$ does not appear it should be assumed that we have set it to unity. We stress that this is the most general flux ansatz consistent with the symmetries of the metric. 

Despite the symmetries of the fluxes mirroring those of the $D=11$ spacetime ansatz, it is known that the existence of $\mathcal{G}$ is inconsistent with supersymmetry in this warped product setting \cite{genLLM}, thus generalising a statement that $\mathcal{G}$ cannot be turned on perturbatively from the LLM class \cite{LLM}.  Therefore, neglecting orbifolds of $AdS_5 \times S^5$ \cite{Aharony:1998xz}, all the known regular\footnote{A well-known remarkable feature of the LLM class of solutions is that supersymmetric solutions are in one-to-one correspondence with solutions to the continuum Toda equation. Despite separable solutions to the Toda existing, such as those of \cite{Spalinski:2005ha}, only one regular solution is known \cite{MN}.   } solutions dual to $D=4$ $\mathcal{N}=2$ SCFTs fit into the LLM class in $D=11$. 

With $\mathcal{G}$ set to zero, one finds the supersymmetry variations as they appear in LLM \cite{LLM}, 
\bea
\g^\mu\partial_\mu\lambda \e_{\pm}
& \mp & \left(\frac1{12}e^{-3\lambda-2A}\g^{\mu\nu} \mathcal{F}_{\mu \nu}
+m\g_5 \right)\e_{\mp}=0,\label{const1}\\
\left(\pm ie^{-A}\g_5+\g^\mu\partial_\mu A\right)\e_{\pm} & \pm &
\left(m\g_5+\frac14 e^{-3\lambda-2A}\g^{\mu\nu} \mathcal{F}_{\mu\nu} \right)\e_{\mp}=0,\label{const2}\\
 \nabla_\mu \e_\pm & \pm & \left( \frac{m}2\g_\mu\g_5+\frac14 e^{-3\lambda-2A} \g^\nu
\mathcal{F}_{\mu \nu}  \right)\e_{\mp}=0.\label{diff3} \eea 

A linear combination of (\ref{const1}) and (\ref{const2}) leads to an algebraic condition independent of $\mathcal{F}$ \be
\left( \g^\mu\partial_\mu (3\lambda+A)\pm
ie^{-A}\g_5\right)\e_\pm \mp
2m\g_5\e_\mp=0. \label{const4} \ee
Further details of various conventions can be found in the appendix. 

In the absence of the four-form flux, $\mathcal{G}$, these equations may be solved by identifying the two spinors \cite{LLM}
\be 
\label{e-ep}
\e_{-} = - \g_5 \e_+,
\ee 
so one only has to solve the Killing spinor equations for a \textit{single} spinor\footnote{One attractive feature of the LLM spinor ansatz is that the vector spinor bilinears $\bar{\e} \g_{\mu} \e, ~\bar{\e} \g_5 \g_{\mu} \e, \bar{\e}^{c} \g_{\mu} \e$ one constructs are all mutually orthogonal and define a natural orthonormal frame.}. In the process one finds a \textit{single} Killing direction. However, when this condition is relaxed, one finds two potentially independent Killing directions \cite{genLLM}, and only when these two directions are identified, does one recover the LLM spinor ansatz (\ref{e-ep}). As the focus of this work is exploring geometries where the two Killing vectors are treated independently, we henceforth relax (\ref{e-ep}) and deal with the various complications. 

Once one relaxes the condition on the spinors one can construct an exhaustive set of scalar and vector bilinears, a complete list of which can be found in appendix A. Furthermore, it can be shown that two of the vectors, $K^1$ and $\Re(K^8)$ in the notation of \cite{genLLM}, are Killing directions and that the warp factors are independent of these directions\footnote{$\Re$ and $\Im$ denote real and imaginary parts.}. We will throughout this work refer to these Killing directions as $X$ and $Y$ respectively. 

Employing a slight rewriting of the results of \cite{genLLM}, we document the following differential conditions on the scalars:
\bea
\label{vs1} e^{A} d(e^{-A} S_1) &=& e^{-A} K^4, \\
\label{vs2} e^{A} d(e^{-A} \Re(T_3)) &=& -e^{-A} \Im(K^7), \\
\label{vs3} e^{-3 \lambda} d(e^{3 \lambda} S_1) &=& 2m \Im(K^8), \\
\label{vs4} e^{-3 \lambda} d(e^{3 \lambda} \Re(T_3)) &=& 2m K^2, \\
\label{vs5} d U_2 &=& - m K^5. 
\eea
%The vectors satisfy similar conditions which, owing to their length, we present in the appendix, (\ref{vec1}) - (\ref{vec10}). 

Furthermore, supersymmetry tells us that the following bilinears are trivial 
\be
U_1 = 
T_1 =  
\Im(S_3) = 0, 
\ee
and that we have two constant scalar bilinears 
\be
d S_2 = d \Im(T_3) = 0. 
\ee
These constants we will refer to henceforth as $s$ and $t$ respectively. As we shall see in due course, it is these two constants that play a pivotal role in determining the final form of the resulting geometry. An extra bonus is that supersymmetry also allows us to determine certain bilinears in terms of these constants: 
\bea
\label{usefulexp}
S_3 &=& - 2m e^{A} s, \nn
T_2 &=& 2 m e^{A} t, \nn
K^3 &=& - s e^{A} d (3 \lambda + A), \nn
\Re(K^7) &=& - t e^{A} d (3 \lambda + A).
\eea
More details on notation and conventions can be found in \cite{genLLM}. 

\section{Killing vectors} 
Recall from \cite{genLLM} that when the Killing vectors $X$ and $Y$ are identified one recovers the LLM geometries \cite{LLM}. The goal of this note is to address the possibility of these two directions being independent. To do this, we will in the next subsection use general techniques based on the spinor bilinears to determine the relationship between these Killing vectors. More particularly, we will determine their inner product, their norms and work out the Lie derivative of one vector with respect to the other. 

Although Killing directions that emerge from the Killing spinor equations typically correspond to R-symmetries, in the subsequent subsection we show that the bilinears and their constituent spinors are indeed generically charged with respect to both $X$ and $Y$. It is in this sense that we label them ``R-symmetries", though as we will appreciate later, only one of them corresponds to a traditional R-symmetry of the internal geometry.  

Having shown that $X$ and $Y$ are R-symmetries, this rules out any immediate connection to the work of Gaiotto and Maldacena \cite{GM}. As a simple check of this, we explicitly show that the $U(1)$ isometry of the hyperbolic space in the Maldacena-N\'u\~{n}ez solution cannot correspond to $X$ or $Y$, or indeed any combination. Therefore, the connection to \cite{GM} is through first recovering LLM by setting $X=Y$ and then inserting a new $U(1)$ isometry by hand.  

\subsection{Commuting vectors}
\label{sec:commute}
Here we determine the inner product of $X$ and $Y$ and also the  Lie derivative of one with respect to the other. We will see in due course that the vectors always commute, a statement that naturally becomes trivial when they are the same vector, i.e. in LLM. A similar calculation with commuting Killing vectors appeared in \cite{Colgain:2011hb}. 

We begin by addressing when the vectors are orthogonal. Using the Fierz identity  (\ref{fierz}), it is possible to show that the inner product of $X$ and $Y$ is 
\be
\label{xdoty}
X \cdot Y = -\frac{2}{3} st \left[1 + 4m^2 e^{2 A} + \frac{e^{2A}}{2} |d (3 \lambda +A) |^2 \right], 
\ee
or by further combining with $K_3 \cdot \Re(K_7)$,  it is possible to directly show that the bracket cannot vanish:
\be
\label{orXY} X \cdot Y = - st  \left[1 + 4m^2 e^{2 A} \right]. 
\ee
In deriving this result we have made use (\ref{usefulexp}). One can confirm that this result is indeed consistent with LLM by employing the following rewritings 
\bea
s = t &=& 1, \nn
2 X = 2 Y &=& - K_{LLM}, \nn
2m e^{A} &=& -  \sinh \zeta_{LLM}, 
\eea
and comparing to (F.31) of LLM \cite{LLM}. These identities allow us also to infer the following useful identity 
\be
\label{iden1}
e^{2A} | d (3 \l +A) |^2 = (1+ 4 m^2 e^{2A}). 
\ee 

Returning to (\ref{orXY}), we observe that the Killing directions $X$ and $Y$ are orthogonal whenever one of $s$ and $t$ are zero, or indeed when both vanish.

We can now calculate the norms of the Killing vectors using (\ref{iden1}) and expressions in the appendix (\ref{contracts1}), (\ref{contracts2}). A straightforward calculation then reveals the following: 
\bea
\label{normX} |X|^2 &=& - \left[ S_1^2 + \Re(T_3)^2  + t^2 (1 + 4 m^2 e^{2A}) \right], \\
\label{normY} |Y|^2 &=& \left[ S_1^2 + \Re(T_3)^2 - s^2 (1 + 4 m^2 e^{2A}) \right] . 
\eea
We now see that $X$ is timelike which is expected since its temporal component cannot be zero without the spinors $\e_{\pm}$ being zero (see appendix). In contrast, whether $Y$ is timelike or spacelike depends on the scalars $S_1$ and $\Re(T_3)$, which are zero for LLM, in which case, (\ref{orXY}), (\ref{normX}) and (\ref{normY}) all agree.  

Now we are on the verge of making an interesting observation. We can show that the norm of the difference $X-Y$ is zero whenever $s=t$. This means that $X-Y$ is either \textit{zero or null}. Then, it is known from the work of \cite{genLLM} that LLM follows once $X$ and $Y$ are identified. This leaves the only unexplored possibility with $s=t$ being a null spacetime. Later, by integrating the differential conditions, we will identify this null spacetime uniquely. 

We now turn attention to whether the vectors $X$ and $Y$ commute or not. This is independent of any choices for $s$ and $t$. Essentially one can ask under what conditions is the Lie derivative of $Y$ with respect to $X$,  ${\cal L}_{X} Y \equiv [X, Y]$, zero. To answer this question, one first determines $d X$ and $d Y$ \cite{genLLM}
\bea
\label{vdiff1} d X &=& \tfrac{m}{2} \left[ \bar{\e}_{+} \g_5 \g_{\mu  \nu} \e_- -  \bar{\e}_{-} \g_5 \g_{\mu  \nu} \e_+ \right] dx^{\mu \nu} + 2 m s e^{-3 \lambda - A}\mathcal{F}, \\
\label{vdiff2} d Y &=& -\tfrac{m}{2} \left[ \bar{\e}_{+} \g_{\mu  \nu} \e_+ -  \bar{\e}_{-} \g_{\mu  \nu} \e_-\right] dx^{\mu \nu} + 2m t e^{-3 \lambda - A} \mathcal{F}.
\eea

The commutator may then be written $i_{X} d Y - i_{Y} d X$. This expression can be divided into a part involving contractions with the two-form flux $\mathcal{F}$ and a part without. We first focus on the part of the commutator involving $\mathcal{F}$.

Using (\ref{usefulexp}) and the expressions \cite{genLLM}
\bea
\label{FX}
i_{X} \mathcal{F} &=& 2 m e^{3 \lambda +2 A} \left[ 2 s  d e^A - K^3 \right] = 2 m s  d \left[  e^{3 \l + 3A}\right], \\
\label{FY} i_{Y} \mathcal{F} &=& 2 m e^{3 \lambda +2 A} \left[ 2 t d e^A - \Re(K^7) \right] = 2 m t  d \left[  e^{3 \l + 3A}\right],
\eea
it is an easy task to convince oneself that these $\mathcal{F}$ dependent terms vanish. 

We now turn attention to the remaining terms. Making use of the following identities:
\bea
\g_{\mu} \g^{\mu \nu} &=& 3 \g^{\nu}, \nn
\g_{\mu} \g^{\rho} \g^{\mu \nu} &=& - \g^{\rho \nu} - 3 \eta^{\rho \nu}, \nn
\g_{\mu} \g^{\rho \sigma} \g^{\mu \nu} &=& - \g^{\rho \sigma \nu} + \g^{\rho} \eta^{\sigma \nu} - \g^{\sigma} \eta^{\rho \nu},
\eea
a lengthy calculation\footnote{At some point it is good to use the identity $\tfrac{1}{2} \bar{\e}_1 \g_{\rho \sigma} \e_2 \bar{\e}_3 \g_5 \g^{\rho \sigma \nu} \e_4 =  \bar{\e}_3 \g_{\rho} \e_4 \bar{\e}_1 \g_5 \g^{\rho \nu} \e_2$. } involving the Fierz identity reveals that
\bea
[X,Y] =  2 m^2 e^{A} \left[ - s \Re(K_7) + t K_3 \right],
\eea
which identically vanishes using (\ref{usefulexp}). As a result $[X,Y] = 0$ without having to make any assumption about the constants $s$ or $t$. In other words, these Killing vectors always commute and the LLM class of geometries where $X=Y$ is just one configuration where this relationship between the vectors becomes trivial. Our hope for the rest of this paper is to identify spacetimes where $X$ and $Y$ are independent. 

\subsection{Killing vectors are R-symmetries}
\label{sec:R}
In this subsection we bring attention to the fact that the spinor bilinears are in general charged with respect to $X$ and $Y$. Some calculations involving the Fierz identity reveal the following relationships: 
\bea
\label{xdotk2} X \cdot K^2 &=& - S_1 s , \\
\label{xdotk4} X \cdot K^4 &=& 2 e^{A} s \Re(T_3), \\
\label{xdotik7} X \cdot \Im(K^7) &=&  2 e^{A} s S_1, \\
\label{xdotik8} X \cdot \Im(K^8) &=&  s \Re(T_3), 
\eea
and
\bea
\label{ydotk2} Y \cdot K^2 &=& - t S_1, \\
\label{ydotk4}Y \cdot K^4 &=& 2 e^{A} t \Re(T_3), \\
\label{ydotik7} Y \cdot \Im(K^7) &=& 2 e^{A} t S_1, \\
\label{ydotik8} Y \cdot \Im(K^8) &=&  t \Re(T_3). 
\eea

Using (\ref{vs1}) - (\ref{vs4})  and the fact that the warp factors are independent of $X$ and $Y$ \cite{genLLM}, we can now infer that the scalar bilinears $S_1$ and $\Re(T_3)$ are charged with respect to both $X$ and $Y$. Note from (\ref{const4}) it is not possible for $\e_+$ to be charged while $\e_-$ is not, since as the warp factors are independent, we have to conclude that both $\e_+$ and $\e_-$ are charged with respect to $X$ and $Y$. 

One can now worry about what happens when one of $s$ and $t$ take special values. When either $s$ or $t$ is set to zero, it would appear that the above bilinears become independent of the $X$ or $Y$ direction. However, by analysing other directions, one can show using (\ref{vs5}) that the spinors are still charged with respect to both
\bea
X \cdot K^5 &=& i t U_2, \nn
Y \cdot K^5 &=&  i s U_2. 
\eea
More generally, for generic $s$, $t$, with $s \neq t$, it can be shown that it is not possible to find a linear combination so that the Killing spinors become independent of this direction.  

However, when $s=t$ we see that the spinors are no longer charged with respect to the direction $X-Y$, and further when $s=t=0$ that the spinors are no longer charged with respect to both $X$ and $Y$. We will deal with these special cases in the next section. For the moment, we stress that the Killing spinors are generically charged with respect to both $X$ and $Y$. We will put this observation to use again in the section \ref{sec:timelike}. 

So far we have assumed that the Killing spinors are completely generic. For example, by confining ourselves to Killing spinors where the scalar bilinear $U_2$ is zero, we see that we can find a linear combination,  i.e. $t X - s Y$, such that the above bilinears are no longer charged with respect to this direction. Then, despite it being a bit awkward, we have the freedom to adopt $X$ and $t X - sY$ as our two Killing directions, so that the Killing spinors are now only charged with respect to one of these directions. However, having set $U_2 =0$, we also have $K^5 = 0$ and using (2.36) of \cite{genLLM}, $K^9 =0$. As we will discuss more fully in the next section, our ansatz generically assumes at least sixteen supersymmetries, a situation which corresponds to the Killing spinor $\e_+$ having only a single component. In general, it can have a maximum of two independent components corresponding to maximal supersymmetry. Therefore, without loss of generality, we can then take $\e_+$ to be a two-component spinor. Then by employing  the explicit gamma matrices in the appendix, it is possible to show that if $U_2 =0$, then $\e_- \propto \g_5 \e_+$, meaning that we are back to LLM. Similar arguments hold when $S_1$ and $\Re(T_3)$ are taken to vanish. 

\subsection{Connection to Gaiotto-Maldacena} 
Since the continuum Toda equation has a reputation for being difficult to solve, starting with the work of Gaiotto-Maldacena \cite{GM} (GM), recent solutions have been constructed by exploiting an extra $U(1)$ symmetry that may be introduced by hand along the Riemann surface. In the process one trades the Toda equation for the cylindrically symmetric Laplace equation and a resulting equivalence with axially symmetric electrostatic problems in three dimensions \cite{Ward:1990qt} (see also \cite{Donos:2010va}). As this $U(1)$ is a global symmetry, while preserving supersymmetry, one can reduce to IIA where it is possible to identify further solutions \cite{ReidEdwards:2010qs, Aharony:2012tz}. 

With this added $U(1)$, we now have solutions with two commuting $U(1)$'s, one corresponding to the original R-symmetry of LLM, and an extra $U(1)$ corresponding to a  global symmetry. While isometries emerging from the Killing spinor equations are typically expected to correspond to R-symmetries, such as the isometries arising in \cite{Gauntlett:2004zh, LLM, Kim:2006qu, Kim:2007hv, Colgain:2010wb, Gabella:2012rc} in the context of $D=11$ supergravity, here we look at a sample geometry in the GM class and confirm that the added $U(1)$ cannot play any role in $X$ or $Y$. 

To do this, we select a prominent example of a spacetime with this extra $U(1)$ symmetry, namely the Maldacena-N\'u\~{n}ez solution \cite{MN}, but with the metric on the hyperbolic space rewritten to highlight the $U(1)$. The overall solution takes the form \cite{Chen:2010jga}
\bea
&&ds^{2} =  \frac{1}{2}
W^{1/3} \biggl[ ds^{2}(AdS_5) + \frac{W^{-1}}{2} \cos^2 \theta ds^2(S^2) + \frac{1}{2} d \theta^2 \nn && \phantom{xxxxxxxxxxxxx} + \frac{1}{2} ds^2(H^2)  +  W^{-1} \sin^2 \theta \left( d \psi + v \right)^2 \biggr], \\ F^{(4)} &=& 
\frac{\cos^2 \theta}{4 W} \biggl[ -\frac{1}{W} [ 3 + \cos^2 \theta ] \sin \theta d \theta (d \psi + v)+  \cos
\theta \vol(H^2) \biggr] \wedge \vol(S^2) \nonumber, 
\eea 
where 
\be
W = 1 + \cos^2 \theta, \quad ds^2(H^2) = 4\left[ \frac{dr^2 + r^2 d \beta^2 }{(1-r^2)^2} \right], \quad  v =  \frac{2 r^2}{1-r^2} d \beta. 
\ee 

Note that this solution corresponds to an analytic continuation of the LLM ansatz, but this distinction will not be important for our purposes. As the solution is expressed in the form of our ansatz (\ref{ansatz}), we can now simply read off the warp factors: 
\be
e^{\lambda} = \frac{W^{1/6}}{\sqrt{2}}, \quad e^{A} = \frac{W^{-1/2} \cos \theta}{\sqrt{2}}.  
\ee

As $X$ and $Y$ both satisfy the same condition (\ref{FX}) and (\ref{FY}) up to the constants $s$ and $t$, we can ask if the Killing vectors $\partial_{\psi}$ and $\partial_{\beta}$ satisfy this relation. While $\partial_{\psi}$ satisfies this condition, a quick calculation reveals that $\partial_{\beta}$ cannot correspond to either $X$ or $Y$ as when contracted into $\mathcal{F}$ it produces a term proportional to $dr$ that cannot be sourced form $d [e^{3 \lambda +3 A}]$. For similar reasons, it cannot correspond to the difference $X-Y$ when $s=t$. 

So we summarise what we have learned in this section. When the Killing spinor ansatz of LLM is generalised, one finds two Killing vectors and the naive expectation is that both of these are ``R-symmetries". By comparing with a typical example of the GM class of geometries, we see that the $U(1)$ of the Riemann surface cannot correspond to either of these Killing directions, thus making a direct connection between our work here and that of GM remote. The connection is then via LLM, since it was shown in \cite{genLLM} that once the Killing vectors are proportional, then they correspond to the same $U(1)$ and the LLM analysis follows. One is then free to insert a global $U(1)$ and recover the work of GM.   

\section{Timelike case}
\label{sec:timelike} 
Now that we have built up a picture of the spinors in terms of spinor bilinears, we can make a statement about the amount of supersymmetry. The warped $S^5 \times S^2$ ansatz already means that we have a large amount of supersymmetry, notably 16 supercharges in the case of LLM \cite{LLM}. Recall from LLM that the spinors $\e_{\pm}$ are assumed to be directly related through (\ref{e-ep}), so there is only a single spinor. Then if one imposes the projection conditions of LLM \cite{LLM} one finds a one component spinor with a phase that depends only on the R-symmetry\footnote{See discussion in \cite{LLM} immediately below (F.48).} . 
 
Now, more generally, we have to incorporate two charges for the spinors, one for each vector $X$ and $Y$. From (\ref{const4}) we have a relationship between the spinors, so they are both charged and we can just focus on one of them. Labeling these vectors as $\partial_{\tau}$ and $\partial_{\psi}$, without loss of generality we can take $\e_+$ to be a four-component spinor with one component carrying a phase of the form $e^{i \tau}$. Now, if we have another vector $\partial_{\psi}$ with respect to which $\e_+$ is also charged, there are only two options. Either this phase multiples the original component of the spinor, in which case it is indistinguishable from the phase in $\tau$, or it is forced to reside in another component of the four-component spinor. Thus, unless $\partial_{\tau}$ and $\partial_{\psi}$ are the same vector, as in LLM, then supersymmetry is automatically doubled as we have one more component in the spinor. In other words, we have to have maximal supersymmetry. 

Then with maximal supersymmetry in $D=11$, it is a well-known theorem \cite{FigueroaO'Farrill:2002ft} that the only solutions are of the Freund-Rubin \cite{Freund:1980xh}, Kowalski-Glikman \cite{KowalskiGlikman:1984wv} or flat spacetime type. There is however a noticeable caveat. We have assumed the spinors are charged to deduce that flat spacetime and the maximally supersymmetric pp-wave are solutions! Indeed, it is precisely for these cases that the spinors are not charged, or only charged under one vector, and this argument does not apply. We will remedy this in subsequent sections. 

Indeed, the conclusion that beyond LLM there are only maximally supersymmetric solutions is already hinted at in (\ref{FX}) and (\ref{FY}). To see this note from the appendix that $X \equiv K^1$ must have a temporal component otherwise the sum of the norms of the spinors becomes zero, so we will assume it is aligned solely along the temporal direction. Then $Y$ becomes spacelike when they are orthogonal. So, we see that if $s=0$ (an extreme $s \neq t$ case), then the flux is not along the temporal direction, i.e. the solution is $AdS_7 \times S^4$, and when $t =0$ the solution is $AdS_4 \times S^7$. Then when both $s = t = 0$, we see that the flux is independent of both the Killing directions and depends on the remaining two transverse directions. This is already suggestively saying that the flux term does not exist and that the solution corresponds to flat spacetime. 

In the rest of this section we address the timelike solutions in turn starting with flat spacetime where our argument does not apply. We thus single out the flat spacetime case where we integrate the differential conditions on the spinor bilinears. Indeed, only in the flat spacetime case does it look manageable to solve for the Killing spinors directly given the spinor bilinears. For $AdS$ spacetimes, we see later by decomposing the Killing spinors from $D=11$ using the gamma matrix decomposition of LLM \cite{LLM, genLLM} that the form of $\e_{\pm}$ is more complicated.  

\subsection{Flat spacetime} 
\label{sec:flat}
In this section we show that flat spacetime follows from the requirement that both of the constants $s$ and $t$ are set to zero. To do this we make use of an explicit decomposition of the gamma matrices in the appendix (\ref{mat}). 

We begin with the facts. $s=t =0$ implies the following additional spinor bilinears are zero through (\ref{usefulexp}): $S_3, T_2, K^3$ and $\Re(K^7)$. From (\ref{xdotk4}) and (\ref{ydotk4}) we also see that $X$, $Y$ and $K^4$ are mutually orthogonal. Also, we note that $X$, $Y$ and $\Im(K^8)$ are also mutually orthogonal from (\ref{xdotik8}) and (\ref{ydotik8}). Then, it is possible to use the Fierz identity to confirm that the inner product of $K^4$ and $\Im(K^8)$ is
\be
K^4 \cdot \Im(K^8) = 2 (t^2 - s^2) m e^{A}, 
\ee
which vanishes for the case at hand. As such, these four vectors define an orthonormal frame, 
so we choose to orient $X$ along $e^0$, $\Im(K^8)$ along $e^1$, $Y$ along $e^2$ and $K^4$ along $e^3$ in accordance with our choice of gamma matrices (\ref{mat}). As this simply amounts to a choice of frame, we are always at liberty to do this. 

Then, noting that the gamma matrices are themselves tensor products, we further decompose the spinors as\footnote{Note only the Killing spinors of flat spacetime can be broken down in such a simple form. }
\be
\e_{\pm} = \theta_{\pm} \otimes \eta_{\pm}. 
\ee
Since we can always rescale $\eta_{\pm}$ relative to $\theta_{\pm}$, without any loss of generality we will take $\theta^{\dagger}_{\pm} \theta_{\pm} = 1$. Plugging these expressions into $K_1^3 = K^3_{2} = K_1^4 = K^4_{2} = 0$, and recalling that $K^4_3$ is, by assumption, non-zero, we arrive at 
\be
{\theta}^{\dagger}_{\pm} \s_{i} \theta_{\pm} = 0, \quad i = 2, 3. 
\ee
Bearing in mind the unit-norm of $\theta_{\pm}$, this leaves overall phases 
\be
\e_+ = \frac{1}{\sqrt{2}} e^{i \beta_{+}} \left( \begin{array}{c} 1 \\ 1 \end{array}\right) \otimes \eta_+, \quad \e_- = \frac{1}{\sqrt{2}} e^{i \beta_{-}} \left( \begin{array}{c} 1 \\ -1 \end{array}\right) \otimes \eta_-,  
\ee
where the final signs in $\theta_{\pm}$ have been set using $S_3 = 0$ with $Y$ non-zero. The overall phases, $\beta_{\pm}$, we can now absorb into $\eta_{\pm}$ through a redefinition. We now turn to determining the form of $\eta_{\pm}$. 

From $s=0$, $K^1 \cdot K^2 = 0$ (via Fierz), $X$ being only along $e^0$ and $K^8$ not being along $e^0$, it is possible to infer the following forms for $\eta_{\pm}$ 
\be
\eta_+ = \left( \begin{array}{c} x_1 e^{i \theta_1} \\ x_2 e^{i \theta_2}  \end{array}\right), \quad \eta_- = \left( \begin{array}{c} x_1 e^{i \theta_3}  \\ x_2 e^{i \theta_4} \end{array}\right),  
\ee
with $x_i, \theta_i \in \mathbb{R}$ and one angular constraint $\theta_1 - \theta_4 = \theta_2 - \theta_3$. Then by ensuring $K^8$ is imaginary along $e^1$, but real along $e^2$ - where it corresponds to $Y$ -  we can narrow down the form of $\eta_{\pm}$ to 
\be
\label{finalspin}
\eta_+ =  e^{i \theta_1} \left( \begin{array}{c} x_1 \\ -i x_2 \end{array}\right), \quad \eta_- = -i e^{i \theta_1}  \left( \begin{array}{c} x_1  \\ i x_2 \end{array}\right). 
\ee

We can now introduce coordinates. 

In this setting, as many scalar bilinears are zero, the norms of $X$ and $Y$ simplify to  
\be
|X|^2 = - |Y|^2 = - \left[ S_1^2 + \Re(T_3)^2 \right]. 
\ee
At this point it is also useful to document the norms of the other two vectors making up the orthonormal frame
\bea
\label{normK4} |K^4|^2 &=& S_1^2 + (t^2-s^2) 4 m^2 e^{2A}, \\
\label{normImK8} |\Im(K^8)|^2 &=& S_1^2 + (t^2 - s^2). 
\eea

Then introducing the vectors $\partial_{\tau}$ and $\partial_{\psi}$ for $X$ and $Y$ respectively, we can determine part of the orthonormal frame. The remaining coordinates come from using (\ref{vs1}) and (\ref{vs3}), and observing that the RHS of these differential equations are closed. Thus, we can determine the orthonormal frame in terms of a single bilinear
\be
e^{0} = (\eta^{\dagger}_+ \s^3 \eta_+ ) d \tau, \quad e^{1} = \frac{e^{-3 \lambda}}{2 (\eta^{\dagger}_+ \s^3 \eta_+ )} dr_5, \quad e^2 = (\eta^{\dagger}_+ \s^3 \eta_{+} )d \psi, \quad e^3 = -\frac{e^{2A}}{(\eta^{\dagger}_+ \s^3 \eta_+ ) } dr_2. 
\ee

We can now integrate (\ref{vs1}) and (\ref{vs3}) to establish that 
\be
r_2 = e^{-A} (\eta^{\dagger}_+ \s^3 \eta_+ ), \quad r_5 = e^{3 \lambda}(\eta^{\dagger}_+ \s^3 \eta_+ ) . 
\ee

Now, by looking at the differential conditions for $X$ (\ref{vdiff1}) and $Y$ (\ref{vdiff2}), we can infer that 
\be
(\eta_+^{\dagger} \eta_+) (\eta^{\dagger}_+ \s^3 \eta_+) = r_5^{-1}, 
\ee
which may be solved in terms of another function $\beta$ 
\be
x_1^2 = r_5^{-1/2} \cosh \beta, \quad x^2_2 = r_5^{-1/2} \sinh \beta. 
\ee
Furthermore, from the fact that the RHS of (\ref{vs2}) and (\ref{vs4}) are closed, we can deduce that $\beta$ is a constant. Finally, if one absorbs various factors involving $\beta$ in the various coordinates and then rescales, 
$r_2 \rightarrow r_2^{-1}$, $r_5 \rightarrow r_5^2$, one recover the usual form of flat spacetime from the $D=11$ spacetime ansatz (\ref{ansatz}),
\be
\label{flatsp}
ds^2 = - d \tau^2 + d \psi^2 + dr_5^2 + r_5^2 ds^2(S^5) + dr_2^2 + r_2^2 ds^2(S^2), 
\ee
where the warp factors and the spacetime $\mathcal{M}_4$ become  
\bea
e^{\lambda} &=& r_5, \quad e^{A} = \frac{r_2}{r_5}, \nn
ds^2(\mathcal{M}_4) &=& \frac{1}{r_5^2} \left[  - dt^2 + dx^2 + dr_5^2 + dr_2^2 \right]. 
\eea
As one final consistency check, one can confirm using  \cite{genLLM}
\bea
\label{vec4} d K^4 &=& \tfrac{m}{2} \left[\bar{\e}_{+} \g_{\mu \nu} \e_{-} + \bar{\e}_{-} \g_{\mu \nu} \e_{+}\right] dx^{\mu \nu} - e^{-3 \lambda -2 A} \Re(T_3) \mathcal{F}, \\
\label{vec7} d \Im(K^7) &=& 
 i \tfrac{m}{2} \left[ \bar{\e}_{+} \g_5 \g_{\mu  \nu} \e_+ +  \bar{\e}_{-} \g_5 \g_{\mu  \nu} \e_-\right] dx^{\mu \nu}- e^{-3 \lambda -2 A} S_1 \mathcal{F}, 
\eea
that indeed $\mathcal{F} = 0$ since despite $S_1$ and $\Re(T_3)$ being non-zero, all other terms in these differential equations are trivial. 

We can also obtain the same result from $D=11$ by simply decomposing the Killing spinor. Since we repeat the process later for $AdS_7 \times S^4$, here we simply state results and omit various details. Note since there is no four-form flux, the Killing spinors are covariantly constant and may be written as 
\be
\eta = e^{- \frac{\a_1}{2} \G_{\a_1 r_5}} \cdots e^{-\frac{\a_5}{2} \G_{\a_5 \a_4}} e^{-\frac{\phi_1}{2} \G_{\phi_1 r_2}} e^{-\frac{\phi_2}{2} \G_{\phi_2 \phi_1}} \eta_0, 
\ee
where $\eta_0$ is a constant spinor and we have parameterised flat spacetime as in (\ref{flatsp}). When one decomposes the spinors in terms of the ansatz of LLM \cite{LLM}
\be
\label{LLMKS}
\eta = \psi \otimes e^{\lambda/2} \left[  \chi_{+} \otimes \e_+ + \chi_- \otimes \e_- \right], 
\ee
one can proceed to read off the components, 
\bea
\label{KSflat}
\e_{+} &=& \frac{1}{2}e^{-\lambda/2} ( 1 - i \g_{3} \g_5 ) \e_0, \nn
\e_{-} &=& \frac{1}{2}e^{-\lambda/2} (\g_{3} \g_5 - i) \e_0, 
\eea
where $\e_0$ is subject to the projector $\g_{1} \g_5 \e_0 = i \e_0$. Observe that, in contrast to $AdS_7 \times S^4$, the simple form of the spinors means that $\e_{+} = \g_{1} \g_5 \e_{-}$. Observe also that this relationship is a direct consequence of (\ref{const1}), when the two-form flux is set to zero, and is expected. Using this condition it is easy to see that $s=t = 0$, another signal that everything is consistent.  

\subsection{$AdS_7 \times S^4$} 
In this subsection we look in detail at the decomposition of the Killing spinors from $D=11$ to extract out the form of $\e_{\pm}$. The final form of the spinors are quite complicated and it is not recommendable to solve the Killing spinor equations this way via spinor bilinears. However, for completeness we determine $\e_{\pm}$ for $AdS_7 \times S^4$ and $AdS_4 \times S^7$. 

The $AdS_7 \times S^4$ solution of $D=11$ supergravity may be expressed as 
\bea
ds^2 &=& ds^2(AdS_7) + \frac{1}{4} ds^2(S^4), \nn
G_4 &=& \frac{3}{8} \vol(S^4),
\eea
where we have adopted the usual normalisations $R_{\mu \nu} = -6 g_{\mu \nu}$ and $R_{mn} = 3 g_{mn}$ for the curvature of $AdS_7$ and $S^4$ respectively. We can now rewrite both $AdS_7$ and $S^4$ as fibrations involving both $S^5$ and $S^2$ respectively  
\bea
ds^2(AdS_7) &=& -\cosh^2 \rho d \tau^2 + d \rho^2 + \sinh^2 \rho d s^2(S^5), \nn
ds^2(S^4) &=& \cos^2 \theta d \psi^2 + d \theta^2 + \sin^2 \theta ds^2(S^2).
\eea
This makes the Killing directions, $\partial_{\tau}$ and $\partial_{\psi}$,  manifest. 

Now we plan to derive the form of $\e_{\pm}$ for $AdS_7 \times S^4$ by taking the known Killing spinors from $D=11$ and decomposing them using the LLM gamma matrices decomposition \cite{LLM}: 
\bea 
\G_{a} &=& \rho_{a} \otimes \s_3 \otimes \g_5, \nn
\G_{\a} &=& 1 \otimes \s_{\a} \otimes \g_5, \nn
\G_{\mu} &=& 1 \otimes 1 \otimes \g_{\mu}, 
\eea
where $a=1,...,5$ denotes $S^5$ directions, $\a=1,2$ denotes $S^2$ directions and $\mu = 0,\dots,3$ labels the remaining directions. Making use of the coordinates $\a_i$, $i=1,...,5$ for $S^5$ and $\phi_{i}$, $i=1,2$ for $S^2$, one can solve the $D=11$ Killing spinor equation 
\be
\label{D11KSE}
\nabla_{M} \eta + \frac{1}{288} \left[ \G_{M}^{~NPQR} - 8 \delta_{M}^{~N} \G^{PQR} \right] G_{NPQR} \eta = 0, 
\ee
leading to the solution 
\bea
\eta &=& e^{- \frac{\rho}{2} \g \G_{\rho}} e^{- \frac{\tau}{2} \g \G_0} e^{-\frac{\a_1}{2} \G_{\a_1 \rho} }  e^{ - \frac{\a_2}{2} \G_{\a_2 \a_1} } \cdots e^{ - \frac{\a_5}{2} \G_{\a_5 \a_4} }  \tilde{\eta}, \nn
\tilde{\eta} &=& e^{- \frac{\theta}{2} \g \G_{\theta}} e^{-\frac{\psi}{2} \g \G_{\psi}} e^{- \frac{\phi_1}{2} \G_{\phi_1 \theta}} e^{-\frac{\phi_2}{2} \G_{\phi_2 \phi_1}} \eta_0, 
\eea
where $\eta_0$ is a constant spinor and we have defined 
\be
\gamma = \G^{\psi \theta \phi_1 \phi_2}. 
\ee

Before proceeding further, it is prudent to keep one eye on the final form of the Killing spinor of LLM (\ref{LLMKS})
%\be
%\label{LLMKS}
%\eta = \psi \otimes e^{\lambda/2} \left[  \chi_{+} \otimes \e_+ + \chi_- \otimes \e_- \right], 
%\ee
where now $e^{\lambda} = \sinh \rho$ is the warp factor, $\psi$ denotes the Killing spinor on $S^5$ and $\chi_{\pm}$ are Killing spinors on $S^2$ satisfying 
\be
\label{s2KS}
\nabla_{\a} \chi_{\pm} = \pm \frac{i}{2} \s_{\a} \chi_{\pm}, \quad \chi_{-} = i \s_3 \chi_+. 
\ee
In terms of our choice of coordinates, these may be expressed as 
\be
\chi_{\pm} = e^{\pm \frac{\phi_1}{2} i \s_1} e^{\frac{\phi_2}{2} i \s_3} \chi^{(0)}_{\pm},  
\ee
where $\chi^{(0)}_{\pm}$ denotes constant spinors. Observe that the second condition in (\ref{s2KS}) means that we cannot have the same constant Killing spinor and instead require $\chi^{(0)}_{-} = i \chi^{(0)}_{+}$. 

To proceed, we decompose the constant spinor as
\be
\eta_0 \equiv \psi_0 \otimes \chi^{(0)}_{+} \otimes \e_0. \ee
Then making use of the relationships 
\bea
\frac{1}{2} (\chi_{+} - i \chi_{-} ) &=& \cos \frac{\phi_1}{2} e^{i \frac{ \phi_2}{2} \s_3} \chi_{+}^{(0)}, \nn
-\frac{i}{2} (\chi_{+} + i \chi_{-} ) &=& \sin \frac{\phi_1}{2} \s_1 e^{i \frac{\phi_2}{2} \s_3} \chi_{+}^{(0)},
\eea
one can determine 
\be
\tilde{\eta} = \psi_0 \otimes \left[ \xi_+ \otimes e^{-i \frac{\theta}{2} \g_{\psi} } e^{i \frac{\psi}{2} \g_{\theta}} \e_0 - i \xi_- \otimes \g_{\theta} \g_{5} e^{i \frac{\theta}{2} \g_{\psi}} e^{i \frac{\psi}{2} \g_{\theta}} \e_0
\right],  
\ee
where we have momentarily redefined 
\be
\xi_{\pm} = \frac{1}{2} (\chi_{+} \mp i \chi_-), ~~\mbox{so that}~~ \s_3 \xi_{\pm} = \pm \xi_{\pm}.   
\ee
To complete the Killing spinor, we impose the projector 
\be
\label{projmax}
\g_{\rho} \g_5 \e_0 = i \e_0, 
\ee
so that 
\bea
\eta &=& \psi \otimes \biggl[ \xi_+ \otimes e^{\frac{\rho}{2} \g_0 \g_5} e^{-i \frac{\theta}{2} \g_{\psi} } e^{i \frac{\psi}{2} \g_{\theta}} e^{i \frac{\tau}{2}} \e_0  \nn &-& i \xi_- \otimes \g_{\theta} \g_{5} e^{\frac{\rho}{2} \g_0 \g_5} e^{i \frac{\theta}{2} \g_{\psi}} e^{i \frac{\psi}{2} \g_{\theta}} e^{i \frac{\tau}{2} } \e_0 \biggr], 
\eea
with $\psi$ now denoting the Killing spinor on $S^5$ satisfying $\nabla_{a} \psi = \frac{i}{2} \rho_{a} \psi$. Imposing the projector (\ref{projmax}) ensures that the Killing spinor fits into the required form of LLM. Observe also that this is one projector less than the number imposed to get the LLM class of solutions (see appendix F of \cite{LLM}) and, as a result, the geometry has maximal supersymmetry, instead of sixteen supersymmetries. Throughout we have defined 
\be
\label{gamma5} 
\g_{5} = i \g_{0 \rho \psi \theta}.  
\ee
Substituting back in for $\chi_{\pm}$ and finally comparing with (\ref{LLMKS}), it is easy to read off the eventual form of the Killing spinors 
\bea
\e_+ &=& \frac{1}{2} e^{-\lambda/2} \left[e^{\frac{\rho}{2} \g_0 \g_5}   e^{-i \frac{\theta}{2} \g_{\psi} } -i \g_{\theta} \g_5 e^{\frac{\rho}{2} \g_0 \g_5} e^{i \frac{\theta}{2} \g_{\psi}}  \right] e^{i \frac{\psi}{2} \g_{\theta}} e^{i \frac{\tau}{2}} \e_0 \nn
\e_- &=& \frac{1}{2} e^{-\lambda/2} \left[ \g_{\theta} \g_5  e^{\frac{\rho}{2} \g_0 \g_5}e^{i \frac{\theta}{2} \g_{\psi}} -i e^{\frac{\rho}{2} \g_0 \g_5}  e^{-i \frac{\theta}{2} \g_{\psi} }   \right] e^{i \frac{\psi}{2} \g_{\theta}} e^{i \frac{\tau}{2}} \e_0. 
\eea
It is possible to see that in agreement with expectations that $s=0$. It is also possible to show that these Killing spinors satisfy the differential conditions for the scalar and vector bilinears presented in generality in \cite{genLLM}.  

\subsection{$AdS_4 \times S^7$} 
The $AdS_4 \times S^7$ solution may be written as
\bea
ds^2 &=& ds^2(AdS_4) + 4 ds^2(S^7), \nn
F^{(4)} &=& 3 \vol(AdS_3), 
\eea
where we have normalised $R_{\mu \nu} = - 3 g_{\mu \nu}$ for $AdS_4$ and $R_{mn} = 6 g_{mn}$ for $S^7$.  Rewriting the solution in terms of the ansatz (\ref{ansatz}), 
\bea
ds^2(AdS_4) &=& - \cosh^2 \rho d \tau^2 + d \rho^2 + \sinh^2 \rho ds^2(S^2), \nn
ds^2(S^7) &=& \cos^2 \theta d \psi^2 + d \theta^2 + \sin^2 \theta ds^2(S^5), 
\eea
the solution to the $D=11$ KSE (\ref{D11KSE}) is 
\bea
\eta &=& e^{- \frac{\rho}{2} \g \G_{\rho}} e^{- \frac{\tau}{2} \g \G_0} e^{-\frac{\phi_1}{2} \G_{\phi_1 \rho} }  e^{ - \frac{\phi_2}{2} \G_{\phi_2 \phi_1} }  \tilde{\eta}, \nn
\tilde{\eta} &=& e^{- \frac{\theta}{2} \g \G_{\theta}} e^{-\frac{\psi}{2} \g \G_{\psi}} e^{- \frac{\a_1}{2} \G_{\a_1 \theta}} \cdots e^{-\frac{\a_5}{2} \G_{\a_5 \a_4}} \eta_0, 
\eea
where $\eta_0$ is a constant spinor and here 
\be
\g = \G^{\tau \rho \phi_1 \phi_2}. 
\ee
In decomposing down to spinors living on $\mathcal{M}_4$ we repeat as before to get 
\bea
\e_+ &=& \frac{1}{2} e^{-\lambda/2} \left[e^{\frac{\theta}{2} \g_{\psi} \g_5}   e^{i \frac{\rho}{2} \g_{0} } -i \g_{\rho} \g_5 e^{\frac{\theta}{2} \g_{\psi} \g_5} e^{-i \frac{\rho}{2} \g_{0}}  \right] e^{i \frac{\tau}{2} \g_{\rho}} e^{i \frac{\psi}{2}} \e_0 \nn
\e_- &=& \frac{1}{2} e^{-\lambda/2} \left[ \g_{\rho} \g_5  e^{\frac{\theta}{2} \g_{\psi} \g_5}e^{-i \frac{\rho}{2} \g_{0}} -i e^{\frac{\theta}{2} \g_{\psi} \g_5}  e^{i \frac{\rho}{2} \g_{0} }   \right] e^{i \frac{\tau}{2} \g_{\rho}} e^{i \frac{\psi}{2}} \e_0, 
\eea
with $\e_0$ a constant spinor satisfying $\g_{\theta} \g_5 \e_0 = i \e_0$ where $\g_5$ is defined in (\ref{gamma5}). Again one can check that the expected algebraic and differential conditions are satisfied and that in this case $t=0$. 

\section{Null case} 
\label{sec:pp}
As noted earlier when $s=t$, we have two options. Either $X=Y$, in which case we return to LLM \cite{genLLM}, or we can take the vector $X-Y$ to be null. At this point significant questions still remain concerning supersymmetry and the expected spacetimes in this class. While on one hand we may expect the Kowalski-Glikman \cite{KowalskiGlikman:1984wv} solution to solve our differential conditions since it fits into the $SO(6) \times SO(3)$ ansatz, on the other hand, as the Killing spinors only depend on the combination $X+Y$, we cannot argue that supersymmetry is maximal and it is possible that there are spacetimes with 16 supersymmetries. Indeed, simply in the class of pp-waves, we can expect solutions other than Kowalski-Glikman,  since it is a well-known fact that all pp-waves preserve at least 16 supersymmetries. 

So to get a better grasp on the geometries in this null class, here we opt to integrate the supersymmetry conditions. In the process we look for other null spacetime solutions which are not pp-waves. As emphasised above, even in the class of pp-wave solutions, there are numerous options in $D=11$ between half-maximal and maximal supersymmetry \cite{Gauntlett:2002cs}. 

\subsection{Kowalski-Glikman} 
From our earlier analysis we know that $X-Y$ is a null vector when $s=t$, so we can introduce the coordinates $x^+,~ x^-$ and associated Killing vectors $\partial_{+}, ~\partial_{-}$,  through defining  
\bea
\label{vectors}
X-Y &=& -C \partial_{+}, \nn
X+Y &=& \mathcal{A} \partial_{-} + \mathcal{B} \partial_{+}, 
\eea
where as $X$ and $Y$ commute, the functions $\mathcal{A}$ and $\mathcal{B}$ are independent of $x^+$, and $C$ is taken to be a constant. 

It is now easy to see from (\ref{vs1}) - (\ref{vs4}) and (\ref{xdotk2}) - (\ref{ydotik8}) that $S_1$ and $\Re(T_3)$ are independent of $X-Y$, so these directions only depend on $x^{-}$. It is also clear from the same equations that we can introduce the following ansatz for $S_1$,  
\be
\label{S1ans}
S_1 = f \cos(\kappa x^{-}), 
\ee
where $f$ now depends only on the, yet to be determined, transverse directions and $\kappa$ is a constant. This in turn determines 
\be
\Re(T_3) = - \frac{\mathcal{A} \kappa f}{4 s} \sin (\kappa x^{-}). 
\ee
Substituting back into (\ref{vs2}), (\ref{xdotik7}) and (\ref{ydotik7}), we can then determine $\mathcal{A}$ to be a constant  
\be
\mathcal{A} = \frac{4 s}{\kappa}, 
\ee
so $S_1$ and $\Re(T_3)$ are now related up to trigonometric functions. For simplicity, we now set $\mathcal{A} =C$ through the choice $\kappa = 4 s C^{-1}$.

In contrast to the timelike flat case, $X, Y, K^4$ and $\Im(K^8)$ are no longer orthogonal, but it is easy to define an orthonormal frame by shifting $K^4$ and $\Im(K^8)$ appropriately, 
\bea
\tilde{K}^{4} &=& K^4 + \frac{2 s e^{A} \Re(T_3)}{S_1^2 + \Re(T_3)^2} (X-Y), \nn 
\Im (\tilde{K}^8) &=& \Im(K^8) + \frac{s \Re(T_3)}{S_1^2 + \Re(T_3)^2} (X-Y). 
\eea
It is easy to check using (\ref{xdotk4}), (\ref{xdotik8}), (\ref{ydotk4}) and (\ref{ydotik8}) that these new directions are orthogonal to the plane spanned by $X$ and $Y$. Observe also that there is no difference between the norm of $K^4$ and $\tilde{K}^4$, or alternatively $\Im(K^8)$ and $\Im(\tilde{K}^8)$, as the shift is along a null direction. 

Using the fact that the RHS of (\ref{vs1}) and (\ref{vs3}) are closed, we can now introduce coordinates, $r_2$ and $r_5$, through defining 
\bea
\label{warpA} e^{-A} f = \frac{1}{r_2}, \\ \label{warpl} e^{3 \lambda} f =  r_5^2.
\eea
In turn, this means that $K^4$ and $\Im(K^8)$ become 
\bea
%K^2 &=& ,\nn
K^{4} &=& -e^{A} f \left[ \cos (\kappa x^-) \frac{dr_2}{r_2} + \kappa \sin(\kappa x^-) dx^- \right] , \nn
\Im(K^8) &=& f \left[  \cos (\kappa x^-) \frac{dr_5}{r_5} - \frac{1}{2} \kappa \sin(\kappa x^-) dx^-  \right].  
\eea

We have chosen the powers in (\ref{warpA}) and (\ref{warpl}) appropriately so that the metric takes a familiar form, though it should be stressed that closure allows us to do this. Indeed, we can go further and choose the form $f$. From $K^{4} \cdot d \lambda$, calculated directly from (\ref{const1}),  we see that $\lambda$ is independent of the coordinate $r_2$. Then by rescaling the coordinate $r_5$ appropriately we can take 
\be
f= \frac{1}{r_5}, \quad e^{A} = \frac{r_2}{r_5}. 
\ee

Note, we can also determine the form of $f$ by calculating the quantity $(X+Y) \cdot d (X-Y)$ as prescribed in appendix B. Observe that when $s=t$ the second term in (\ref{XYrel}) disappears and if one follows the calculation through (with $m=1$), one finds that $f = r_5^{-1}$. 

Now, using the norms (\ref{normK4}) and (\ref{normImK8}) we can determine $g^{r_2 r_2}$ and $g^{r_5 r_5}$, components of the inverse metric. Similarly from (\ref{normX}) and (\ref{normY}) we can work out the rest of the metric. The form of the spacetime then becomes 
\bea
ds^2(\mathcal{M}_4) &&=   \frac{4 }{C^2 r_5^2} \left[ dx^+ dx^-  - \left( s^2 (r_5^2 + 4 r_2^2)  + \frac{\mathcal{B}}{C}\right) (dx^-)^2 \right] \nn && \phantom{xxxxxxxxxxxxx} + \frac{1}{r_5^2} ( dr_2^2 + dr_5^2). 
\eea

At this point it is worth observing that the overall $D=11$ spacetime is of the form of a pp-wave since $e^{\lambda} = r_5$. It is also looking suggestive that the pp-wave in question may be the maximally supersymmetric pp-wave. To confirm this we can determine the flux term $\mathcal{F}$ by contracting in the vector $X+Y$ and using (\ref{FX}) and (\ref{FY}). The two-form flux term of the $D=11$ ansatz (\ref{ansatz}) may be determined to be of the form 
\be
\mathcal{F} = \frac{12}{ C} s r_2^2 dx^- \wedge dr_2. 
\ee

The Kowalski-Glikman solution (\ref{KGmet}) then simply corresponds to the choice\footnote{Here we have set $\mathcal{B} =0$ by hand. However, more generally if it is assumed to be of the usual form $\mathcal{B} = B_{ij} x^i x^j$ in Cartesian coordinates, where $B_{ij}$ is a real constant symmetric matrix, (19) of \cite{Gauntlett:2002cs} tells us that for greater than sixteen supersymmetries we require $\sum_j B_{ij} \G^j \chi = 0$. Therefore, we either have maximal supersymmetry when $\mathcal{B} =0$, or half the supersymmetry when $\mathcal{B} \neq0$.}
\be
C = \sqrt{2}, \quad s = \frac{\mu}{6 \sqrt{2}}, \quad \mathcal{B} = 0. 
\ee
%where $\mathcal{B} = 0$ follows from the Einstein equation, an equation that is not implied by supersymmetry \cite{Gauntlett:2003wb}. More precisely, the Einstein equation tells us that $\partial_i \partial^i \mathcal{B} = 0$, which means that for the transverse space $\mathbb{R}^9$ that $\mathcal{B}$ is linear in the coordinates of $\mathbb{R}^9$.  This is then inconsistent with the assumed $SO(6) \times SO(3)$ symmetry of the LLM ansatz. 

So, in summary, we have identified the null spacetime that arises from the LLM ansatz uniquely. Not only is it a pp-wave, but it is the maximally supersymmetric pp-wave, or Kowalski-Glikman solution \cite{KowalskiGlikman:1984wv}. For completeness we decompose the Killing spinors from $D=11$ in the next subsection to work out the Killing spinors $\e_{\pm}$. 

\subsection{Decomposition}
Here we derive the Killing spinor for the Kowalski-Glikman solution \cite{KowalskiGlikman:1984wv}, using spherical coordinates to make the $SO(6) \times SO(3)$ isometry manifest. Essentially our analysis parallels that of \cite{FigueroaO'Farrill:2001nz} modulo this change in coordinates. The KG solution may be written as
\bea
\label{KGmet}
ds^2 &=& 2 dx^+ dx^- - \frac{\mu^2}{36} \left( 4 r_2^2 + r_5^2 \right) (dx^-)^2 + dr_2^2 + r_2^2 ds^2(S^2) + d r_5^2 + r_5^2 ds^2(S^5), \nn
F^{(4)} &=& \mu r_2^2 dx^- \wedge dr_2 \wedge \vol(S^2). 
\eea
We introduce the vielbein 
\bea
e^{-} &=& dx^-, \nn
e^{+} &=& dx^+ - \frac{\mu^2}{72} (4 r_2^2 + r_5^2) dx^-, \nn
e^{1} &=& dr_2, \quad e^{2} = r_2 d \phi_1, \quad e^{3} = r_2 \sin \phi_1 d \phi_2, \nn 
e^{4} &=& dr_5, \quad e^{5} = r_5 d \a_1, \dots, e^{9} = r_5 \sin {\a_1} \dots \sin{\a_4} d \a_5, 
\eea
where we have used the usual nested expressions for coordinates on the respective spheres. 

Writing the Killing spinor equation as 
\be
\nabla_{M} \eta + \Omega_{M} \eta = 0, 
\ee
expressions for the various $\Omega_{M}$ take the form 
\bea
\Omega_{+} &=& 0, \nn
\Omega_{-} &=& - \frac{\mu}{12} \left( \G_+ \G_- +1 \right) I ,\nn
\Omega_{i} &=& \frac{\mu}{6} \G_+ I \G_{i} ,~~i =1,2,3, \nn
\Omega_{i} &=& \frac{\mu}{12} \G_+ I \G_{i}, ~~i=4,\dots,9,   
\eea
where $I = \G_{123}$. 

It is easy to solve for the standard Killing spinors, namely those in the kernel of $\G_{+}$, with the only difference in moving from cartesian to spherical coordinates being that one has more spin connection expressions which enter the analysis. As always the Killing spinors are independent of $x^+$ as both $\Omega_{+}$ and the spin connection in this direction vanish. The standard Killing spinors are thus 
\bea
\eta_{\textrm{stan}} &=& e^{\frac{\mu}{4} I x^-} e^{-\frac{\phi_1}{2} \G_{2 1}} e^{-\frac{\phi_2}{2} \G_{3 2}} e^{-\frac{\a_1}{2} \G_{54}} \cdots e^{-\frac{\a_5}{2} \G_{98}} \psi_{+}, 
\eea
where $\psi_+$ is a constant spinor satisfying $\G_{+} \psi_+ = 0$. Up the the presence of the $x^-$ dependence, this is just the Killing spinor equation in flat spacetime. This is entirely expected as the flux terms all come with an $x^-$ component, which when lowered, kills $\psi_+$. 

Our task now is to find the remaining supernumerary Killing spinors, a task that is complicated somewhat by the $\Omega_i$ terms not immediately vanishing. Using $\Omega_i \Omega_j = 0$ it is possible to show that the supernumerary Killing spinors are linear in just $r_2$ and $r_5$ with the final form of the $D=11$ Killing spinor being 
\bea
\eta &=& (1 - r_2 \Omega_1 - r_5 \Omega_4) \left[ e^{\frac{\mu}{4} I x^-} \tilde{\psi}_+ + e^{\frac{\mu}{12} I x^-}\tilde{\psi}_- \right] , 
\eea
where in $\tilde{\psi}_{\pm}$ we have absorbed all the angular dependence. These spinors are subject to the projectors $\G_{\pm} \tilde{\psi}_{\pm} = 0$. 

Then, proceeding as before, we can decompose the $D=11$ Killing spinor and extract out $\e_{\pm}$ from the LLM ansatz: 
\bea
\e_+ &=& \frac{1}{2} e^{-\lambda/2} \biggl[ (1-i \g_{r_2} \g_5 ) e^{i \frac{\mu}{4} \g_{r_2} x^-} \psi_+ +  (1-i \g_{r_2} \g_{5} ) e^{i \frac{\mu}{12} \g_{r_2} x^-} \psi_- \nn &-& i r_2 \frac{\mu}{6} \g_+ (1+ i \g_{r_2} \g_5) e^{i \frac{\mu}{12} \g_{r_2} x^-} \psi_- - i r_5 \frac{\mu}{12} \g_+ (1- i \g_{r_2} \g_5) e^{i \frac{\mu}{12} \g_{r_2} x^-} \psi_- \biggr], \nn
\e_- &=& \frac{1}{2} e^{-\lambda/2} \biggl[ (\g_{r_2} \g_5 -i) e^{i \frac{\mu}{4} \g_{r_2} x^-} \psi_+ +  ( \g_{r_2} \g_{5} -i ) e^{i \frac{\mu}{12} \g_{r_2} x^-} \psi_- \nn 
&+& i r_2 \frac{\mu}{6} \g_+ (\g_{r_2} \g_5+i ) e^{i \frac{\mu}{12} \g_{r_2} x^-} \psi_- + i r_5 \frac{\mu}{12} \g_+ (\g_{r_2} \g_5 -i) e^{i \frac{\mu}{12} \g_{r_2} x^-} \psi_-
\biggr]. 
\eea
Here in addition to $\G_{\pm} \psi_{\pm} = 0$, the constant spinors $\psi_{\pm}$  also satisfy $\g_{r_5} \g_5 \psi_{\pm} = i \psi_{\pm}$. Notice here that when we take $\mu \rightarrow 0$ we recover the Killing spinors for flat spacetime (\ref{KSflat}), so we can have some degree of confidence in this result. Going further one can determine the form of the two constants $s, t$. Once the constant spinors are scaled correctly, in terms of the mass parameter $\mu$ they take the form 
\be
s = t = \mu, 
\ee
thus giving a physical meaning to these constants. Again, using the explicit form of the spinors, one can check that the expected geometric conditions on the bilinears are satisfied. 

\section{Discussion} 
In this work we have attempted to address a notable loose-end in the analysis of \cite{genLLM}. While this earlier work did rule out the existence of an additional flux term in the context of the earlier LLM ansatz \cite{LLM}, when one generalised the Killing spinor ansatz, two Killing vectors were found. As geometries dual to $\mathcal{N} =2$ SCFTs only realise $SU(2) \times U(1)$ R-symmetry, this raised a pertinent question about the nature of the mysterious second isometry direction. Thus, the goal of this paper was to identify solutions beyond LLM in which these two Killing directions are manifest. 

So, in this paper we have performed a further study of these two Killing directions. By using general techniques we have shown that they always commute and that their inner product is proportional to the product of two fundamental scalar bilinears. In addition, we have confirmed expectations that the isometry directions correspond to R-symmetries and shown that the connection to the work of Gaiotto-Maldacena is through LLM. We have then argued that the presence of two independent R-symmetries means that supersymmetry will be enhanced beyond the sixteen supersymmetries of LLM leading to geometries with maximum supersymmetry. Where this argument fails to hold, namely for flat spacetime and pp-waves, we have integrated the supersymmetry conditions directly. 

Interestingly, through our work here, we see that constant scalar bilinears play a central role in determining the final form of the geometry. While these scalars are typically normalised to unity, we observe that when one of the two is set to zero we find an $AdS$ spacetime, whereas if both are set to zero, flat spacetime is the only outcome. Moreover, when both constants are equal and there is a null Killing vector, we have shown that these constants correspond to the mass parameter in the Kowalski-Glikman solution. Therefore, our overarching description allows us to put all the maximally supersymmetric solutions on the same footing as LLM. However, we must caution that supersymmetric interpolating flows from $AdS_7 \times S^4$ to LLM,  generalising those based on dimensional reduction \cite{Anderson:2011cz}, are not expected as the end-points correspond to different values of the fundamental constants. In other words, if interpolating solutions exist, we can expect them to be non-supersymmetric, or to not include one of the end-points. 

Echoing the introduction, we stress that the identification of the isometries with those of the maximally supersymmetric geometries brings the more general form of the Killing spinor ansatz of LLM to its logical conclusion. A similar outcome was noted in \cite{Colgain:2011hb} where, an LLM-type ansatz in type IIB proved to be inconsistent with the existence of a $U(1)$ R-symmetry, and in the process, $AdS_5 \times S^5$ was recovered.  As such, the $SU(2)$ should be non-geometric for $\mathcal{N} = 2$ SCFT duals in type IIB. The work here suggests that new geometries dual to $\mathcal{N}=2$ SCFTs may be found in type IIB supergravity by searching for an extra $U(1)$ in the classification of \cite{Gauntlett:2005ww}. 
It remains to be seen if any of them are regular.

\subsection*{Acknowledgements}

We are grateful to P. Galli, J. B. Wu and H. Yavartanoo  for early collaboration. In particular, we thank H. Yavartanoo for useful and timely insights. In addition, we would like to thank Y. Lozano, J. Figueroa O'Farrill and K. Sfetsos for discussion. We would like to acknowledge the hospitality of the CERN theory group where this draft came together. This research is partially supported by the research grants MICINN-09-FPA2009-07122 and MEC-DGI-CSD2007-00042.

\appendix
\section{Conventions}
We will be borrowing our conventions from \cite{LLM} where we refer the reader for details. Here we provide a brief summary. On the external space $\mathcal{M}_4$ with signature $(-,+,+,+)$ we will take $\e^{0123} = 1$. As a result, defining  $\g_5=i\g_{0123}$, we have $ \g_{5}^2 = +1$. We will then adopt
\bea
\label{pc}
(\g_0)^\dagger=-\g_0, \quad (\g_i)^\dagger=\g_i, \eea
Also from  LLM \cite{LLM}, we see that the intertwiners $A$ and $C$ are given by
\be
A \equiv \g^{0}, \quad C \equiv \g^2.
\ee
From (\ref{pc}) this means that
\be
A \g_{\mu} A^{-1} = - \g_{\mu}^{\dagger}. \quad
\ee
In LLM $\g^2$ is antisymmetric and $\g^{0}, \g^{1}, \g^{3}$ symmetric so that
\be
C^{-1} \g_{\mu}^{T} C = - \g_{\mu}.
\ee
Note that subject to these choices
\be
\g_5^{\dagger} = \g_5, \quad \g_5^{T} = - \g_5.
\ee

Then defining $D = C A^{T}$ in the usual fashion, we can define the conjugate spinor to $\e$ as $\e^{c} = D \e^{*} = \g^{2} \g^{0} \e^{*}$. This implies that $\bar{\e}^{c} = - \e^{T} \g^{2}$. Note also that $D= \g^{2} \g^{0}$ and that $D D^* = +1$, so that we have the freedom to take $\e$ to be a Majorana spinor provided we impose $\e^{c} = \e$.

Given spinors $\chi, \xi$ and spinor bilinears constructed from $p$ antisymmetrised gamma matrices $\g^{(p)} \equiv \g^{\mu_1 \cdots \mu_p}$, we have the following symmetry properties for the spinor bilinears
\bea
(\bar{\chi} \g^{(p)} \xi)^{\dagger} &=& (-1)^{\tfrac{p(p+1)}{2} +1} \bar{\xi}  \g^{(p)} \chi,  \nn
(\bar{\chi}^c \g^{(p)} \xi)^{T} &=& (-1)^{\tfrac{p(p+1)}{2} +1} \bar{\xi}^c  \g^{(p)} \chi.
\eea

Finally, an explicit representation of the above $\gamma$ matrices may be written:
\bea
\label{mat}
\g_0 &=& 1 \otimes i \sigma_3, \quad
\g_1 = 1 \otimes \sigma_1, \quad
\g_2 = \sigma_1 \otimes \sigma_2, \quad
\g_3 = \sigma_2 \otimes \sigma_2.
\eea
Employing this choice $\g_5 \equiv i \g_{0123} = \sigma_3 \otimes \sigma_2$.

The $D=4$ Fierz identity is 
\bea
\label{fierz}
\bar{\e}_1 \e_2 \bar{\e}_3 \e_4 = \frac{1}{4} \biggl[  \bar{\e}_3 \e_2 \bar{\e}_1 \e_4 &+& \bar{\e}_3 \g_5 \e_2 \bar{\e}_1 \g_5 \e_4 + \bar{\e}_3 \g_{\rho} \e_2 \bar{\e}_1  \g^{\rho} \e_4 \nn &-& \bar{\e}_3 \g_5 \g_{\rho} \e_2 \bar{\e}_1 \g_5 \g^{\rho} \e_4 - \frac{1}{2} \bar{\e}_3 \g_{\rho \sigma} \e_2 \bar{\e}_1 \g^{\rho \sigma} \e_4\biggr]. 
\eea

\subsection{Bilinear zoo} 
\bea S_1&=&\frac{i}2(\bar{\e}_+\e_++\bar{\e}_-\e_-), \quad S_2=\frac{i}2(\bar{\e}_+\e_+-\bar{\e}_-\e_-), \quad S_3=\bar{\e}_+\e_-, \nn T_1&=&\frac12(\bar\epsilon_+\g_5\epsilon_++\bar\epsilon_-\g_5\epsilon_-), \quad
T_2=\frac12(\bar\epsilon_+\g_5\epsilon_+-\bar\epsilon_-\g_5\epsilon_-), \quad
T_3=\bar\epsilon_+\g_5\epsilon_-, \nn
U_1 &=& \bar{\e}_{+}^{c} \e_{-}, \quad U_2 = \bar{\e}_{+}^{c} \g_5 \e_-,
\eea

\bea
K^{1}_{\mu} &=& \tfrac{1}{2} ( \bar{\e}_+ \g_{\mu} \e_+ + \bar{\e}_- \g_{\mu} \e_- ), \quad K^{2}_{\mu} = \tfrac{1}{2} ( \bar{\e}_+ \g_{\mu} \e_+ - \bar{\e}_- \g_{\mu} \e_- ) \nn K^{3}_{\mu} &=& \tfrac{1}{2} ( \bar{\e}_+ \g_5 \g_{\mu} \e_+ + \bar{\e}_- \g_5 \g_{\mu} \e_- ), \quad K^{4}_{\mu} = \tfrac{1}{2} ( \bar{\e}_+\g_5 \g_{\mu} \e_+ - \bar{\e}_- \g_5 \g_{\mu} \e_- ), \nn
K^{5}_{\mu} &=& \tfrac{1}{2} ( \bar{\e}_+^c \g_{\mu} \e_+ + \bar{\e}_-^c \g_{\mu} \e_- ), \quad K^{6}_{\mu} = \tfrac{1}{2} ( \bar{\e}_+^c \g_{\mu} \e_+ - \bar{\e}_-^c \g_{\mu} \e_- ), \nn
K^7_{\mu} &=& \bar{\e}_+ \g_{\mu} \e_-, \quad K^8_{\mu} = \bar{\e}_+ \g_5 \g_{\mu} \e_-, \nn
K^9_{\mu} &=& \bar{\e}_+^c \g_{\mu} \e_-, \quad K^{10}_{\mu} = \bar{\e}_+^c \g_5 \g_{\mu} \e_-.
\eea

\subsection{Useful stuff} 
Here we record some relationships derived via Fierz identity. They relate to contractions between vector bilinears:
\bea
\label{contracts1}
\bar{\e}_{+} \g_{\mu} \e_+ \bar{\e}_{+} \g^{\mu} \e_+ &=& (\bar{\e}_+ \e_+)^2 - (\bar{\e}_+ \g_5 \e_+)^2 = - \bar{\e}_+ \g_5 \g_{\mu} \e_+ \bar{\e}_+ \g_5 \g^{\mu} \e_+, \nn
\bar{\e}_{-} \g_{\mu} \e_- \bar{\e}_{-} \g^{\mu} \e_- &=& (\bar{\e}_- \e_-)^2 - (\bar{\e}_- \g_5 \e_-)^2 = - \bar{\e}_- \g_5 \g_{\mu} \e_- \bar{\e}_- \g_5 \g^{\mu} \e_-, \nn
\bar{\e}_{+} \g_{\mu} \e_- \bar{\e}_{+} \g^{\mu} \e_- &=& (\bar{\e}_+ \e_-)^2 - (\bar{\e}_+ \g_5 \e_-)^2 = - \bar{\e}_+ \g_5 \g_{\mu} \e_- \bar{\e}_+ \g_5 \g^{\mu} \e_-, \nn
\bar{\e}_{-} \g_{\mu} \e_+ \bar{\e}_{-} \g^{\mu} \e_+ &=& (\bar{\e}_- \e_+)^2 - (\bar{\e}_- \g_5 \e_+)^2 = - \bar{\e}_- \g_5 \g_{\mu} \e_+ \bar{\e}_- \g_5 \g^{\mu} \e_+,
\eea
\bea
\label{contracts2}
\frac{1}{2} \left( \bar{\e}_{+} \g_{\mu} \e_+ \bar{\e}_{-} \g^{\mu} \e_- -  \bar{\e}_+ \g_5 \g_{\mu} \e_+ \bar{\e}_- \g_5 \g^{\mu} \e_- \right) &=& \bar{\e}_- \e_+ \bar{\e}_+ \e_- -   \bar{\e}_- \g_5 \e_+ \bar{\e}_+ \g_5 \e_-, \nn
\frac{1}{2} \left( \bar{\e}_{+} \g_{\mu} \e_- \bar{\e}_{-} \g^{\mu} \e_+ -  \bar{\e}_+ \g_5 \g_{\mu} \e_- \bar{\e}_- \g_5 \g^{\mu} \e_+ \right) &=& \bar{\e}_- \e_- \bar{\e}_+ \e_+ -   \bar{\e}_- \g_5 \e_- \bar{\e}_+ \g_5 \e_+. 
\eea

\section{Details of Fierz calculations} 
In this section we give some details of how to calculate the following 
\be
\label{inprod}
( X + Y ) \cdot (d X - d Y). 
\ee
First note that in the text we have already stated that $X \cdot d Y - Y \cdot d X$, i.e. that the vectors commute. This result we got by similar methods to below, so accepting this result, (\ref{inprod}) reduces to calculating 
\be
X \cdot d X - Y \cdot d Y. 
\ee
Using (\ref{vdiff1}), (\ref{vdiff2}) and the Fierz identity, after a number of cancellations, it is possible to show that 
\bea
&& (X \cdot d X - Y \cdot d Y)^{\nu} = \frac{m}{8} \biggl[ - 24 \left( S_1 \Im(K^8) + \Re(T_3) K^2 \right)^{\nu} \nn && \phantom{xx} 
-4 K^4_{\rho} \left( \bar{\e}_+ \g^{\rho \nu} \e_- + \bar{\e}_- \g^{\rho \nu} \e_+ \right) - 4 i  \Im(K^7)_{\rho} \left( \bar{\e}_+ \g_5 \g^{\rho \nu} \e_+ + \bar{\e}_- \g_5 \g^{\rho \nu} \e_- \right) 
\biggr] .  
\eea
Applying a second round of Fierz identities to the lower line, we can replace the contractions 
\bea
&& 4 K^4_{\rho} \left( \bar{\e}_+ \g^{\rho \nu} \e_- + \bar{\e}_- \g^{\rho \nu} \e_+ \right) + 4 i  \Im(K^7)_{\rho} \left( \bar{\e}_+ \g_5 \g^{\rho \nu} \e_+ + \bar{\e}_- \g_5 \g^{\rho \nu} \e_- \right)  \nn && = -\frac{2}{m} (X \cdot dX - Y \cdot d Y)^{\nu} -12 (S_3 K^3 + T_2 \Re(K^7))^{\nu}, 
\eea
leading to 
\bea
\label{XYrel}
&& (X \cdot d X - Y \cdot d Y)^{\nu} = \frac{m}{6} \biggl[ - 24 \left( S_1 \Im(K^8) + \Re(T_3) K^2 \right)^{\nu} \nn
&& \phantom{xxxxxxxx} + 24 m e^{2A} (s^2-t^2) (d (3 \lambda +A) )^{\nu} \biggr]. 
\eea
Here we have used (\ref{usefulexp}) to rewrite the last expression.

\end{document}